\documentclass[12pt]{article}

\usepackage{cite}
\catcode`@=11 \@addtoreset{equation}{section} \catcode`@=12

\begin{document}
\begin{titlepage}
\vfill
\begin{center}
{\Large \bf Effective $SO$ Superpotential for ${\cal N}=1$  Theory
with $N_f$ Fundamental Matter}\\[1cm]
Pravina Borhade \footnote{E-mail: pravina@phy.iitb.ac.in},
P. Ramadevi\footnote{Email: ramadevi@phy.iitb.ac.in}\\
{\em Department of Physics, \\Indian Institute of Technology Bombay,\\
Mumbai 400 076, India}\\
\end{center}
\vfill
\begin{abstract}


Motivated by the duality conjecture of
Dijkgraaf and Vafa between supersymmetric gauge
theories and matrix models, we derive the effective
superpotential of ${\cal N}=1$ supersymmetric gauge theory 
with gauge group $SO(N_c)$ and arbitrary tree level polynomial superpotential 
of one chiral superfield in the adjoint representation and $N_f$ fundamental
matter multiplets. For a special point in the 
classical vacuum where the gauge group is 
unbroken, we  show that the effective superpotential matches with
that obtained from the geometric engineering approach. 

\end{abstract}
\vfill
\end{titlepage}



\section{Introduction}
Large $N$ topological duality relating $U(N)$ Chern-Simons gauge theory 
on $S^3$ to $A$-model topological string \cite{gv1} and its embedding 
in the superstring context \cite{vafa1} has led to interesting 
interconnections between geometry of Calabi-Yau 
three-folds ($CY_3$) and ${\cal N}=1$ supersymmetric gauge theories.
Strong coupling dynamics of supersymmetric gauge theories
can be studied within the superstring duality \cite{vafa1} by
geometrically engineering $D$-branes. Using the
geometric considerations of dualities in IIB string theory,
Cachazo et al \cite{civ1} have obtained low energy effective 
superpotential for a class of $CY_3$ geometries whose
singular limit is given by
\begin{equation}
W'(x)^2+y^2+z^2+w^2=0~, \label {singu}
\end{equation}
where $W(x)$ is a polynomial of degree $n+1$. 
In fact, the low energy effective superpotential corresponds to a  
${\cal N}=1$ supersymmetric $U(N)$ Yang-Mills with adjoint scalar $\Phi$ 
and tree level superpotential $W_{tree}(\Phi)=\sum_{k=1}^{n+1} (g_k/k)
Tr \Phi^k$.

The mirror version of the large $N$ topological duality conjecture\cite{gv1}
was considered in ref. \cite{dv1} relating topological $B$ strings
on the $CY_3$ geometries \cite{civ1} to matrix models. The potential of the 
matrix model $W(\Phi)=(1 / g_s) W_{tree}(\Phi)$  where $\Phi$ 
denotes a hermitian matrix.  Further, Dijkgraaf-Vafa have 
conjectured that the low-energy effective superpotential can be 
obtained from the planar limit of these matrix models \cite{dv1,dv2,dv3}. 
The Dijkgraaf-Vafa conjecture was later on proved by various methods: 
(i) by factorization of Seiberg-Witten curves \cite {ferr1}, 
(ii) using perturbative field theory arguments 
\cite {dglvz1} and (iii) generalized Konishi anomaly approach \cite {cdsw1}.

The extension of topological string duality relating
Chern-Simons theory with $SO/Sp$ gauge groups to $A$-model
closed string on an orientifold of the resolved conifold
was studied by Sinha-Vafa \cite {sv1}.
Generalizing the geometric procedure considered for $U(N)$ \cite {civ1}, 
the effective superpotential for ${\cal N}=1$ supersymmetric 
theories with $SO/Sp$ gauge groups with 
$W_{tree}= \sum_k (g_{2k}/  2k) Tr \Phi^{2k}$ where $\Phi$ is
adjoint scalar superfield  were derived for the
orientifolds of the $CY_3$ geometries \cite {eot1}. 
These effective superpotentials have also been
computed within perturbative gauge theory \cite {iho1},
using  matrix model techniques in \cite {achkr1} and using
the factorization property of ${\cal N}=2$ Seiberg-Witten curves
\cite {jo1}.

Related works involving second rank tensor matter fields
have been considered in refs.\cite{ikrsv1,argurio1,ac1}.
The geometric engineering of ${\cal N}=1$ gauge theories with unitary
gauge group and matter in the adjoint and symmetric or antisymmetric
representations has been investigated in \cite{llt1}.
Also for the $SO/Sp$ theory with symmetric/antisymmetric tensor,
the geometric construction was studied in \cite{llt3}.

So far, the effective superpotential computation involved
${\cal N}=1$ supersymmetric gauge theories with either adjoint matter
or second rank tensor matter. 
The inclusion of matter transforming in the fundamental representation 
of these gauge groups can also be studied within the Dijkgraaf-Vafa setup
\cite{acfh1,mcg1,br1,suzuki1,reino1,ow1,fo1}.
For $U(N)$ gauge group, it was shown that the effective superpotential 
gets contributions from  the matrix model planar diagrams with
zero or one boundary \cite {acfh1,br1}. 
In \cite {reino1}, it has been shown that 
the $U(N)$ effective superpotential for
theories with $N_f$ fundamental flavors can be calculated in terms of
quantities computed in the pure gauge theory.
Chiral ${\cal N}=1$ $U(N)$ gauge theories with antisymmetric,
conjugate symmetric, adjoint and fundamental matter have also been
studied in Refs.\cite {llt2, llt3}.


Further, geometric engineering of the supersymmetric theories 
with $N_f$ fundamental flavors was considered in \cite {ook1,rtw1} by 
placing $D5$ branes at locations given by the the mass 
$m_a$ ($a=1,2,\ldots N_f$) which are not the zeros of $W'(x)=0$. 
The $SO/Sp$ effective superpotential computations from 
matrix model approach have been presented for tree level 
superpotential of adjoint matter upto quartic terms \cite {ow1,fo1}.
Though there are several indications as to how the 
$U(N)$ effective superpotential can be related to $SO/Sp$ groups, it 
is certainly not a proof. So we need to explicitly obtain the results
using an independent method as elaborated in this paper.
We consider ${\cal N}=1$ supersymmetric $SO(N_c)$ gauge theory
with arbitrary tree level superpotential of one chiral superfield in the
adjoint representation and $N_f$ fundamental matter multiplets.
We use the technique developed in \cite {reino1} to calculate the 
effective superpotential of this theory.

The organization of the paper as follows: In section 2, we briefly
discuss the relevant matrix model and its free energy. Then we
discuss the $SO(N_c)$ effective superpotential for ${\cal N}=1$
supersymmetric theory with fundamental matter in section 3.
In particular, for a point in classical vacuum
where the gauge group is unbroken, we obtain a neat expression for the
effective superpotential for a most general tree level superpotential 
involving one adjoint matter field. We also give a formal 
expression for a generic vacua where the 
gauge group is broken. In section 4, we recapitulate the geometric 
considerations of dualities. Then, we  evaluate the effective superpotential 
for an example involving sixth power tree level potential 
and show that the results agree with our expressions in section 3 for the 
unbroken case. We conclude with summary and discussions in section 5.

\section{Relevant Matrix Model}
Let us consider ${\cal N}=1$ supersymmetric $SO(N_c)$ gauge theory with one
adjoint field $\Phi$ and $N_f$
flavors of quarks $Q^I$'s with mass $m_I$'s ($I=1,2,\ldots N_f$) in the 
vector(fundamental) representation. The tree level superpotential of this
theory is given by \cite{ow1}
\begin{equation}
W_{tree}=W(\Phi)+\sum_{I=1}^{N_f}\left(\tilde{Q} \Phi Q + 
m_I \tilde{Q}Q\right)~, 
\end{equation} 
where $W(\Phi)$ is a polynomial with even powers of $\Phi$:
\begin{equation} 
W(\Phi) = \sum_{k=1}^{n+1} \frac{g_{2k}}{2k} Tr{\Phi}^{2k} \label{tree1}\, .
\end{equation}
According to Dijkgraaf-Vafa conjecture, the effective superpotential of
this theory can be obtained from the planar limit of the
matrix model whose tree level potential
is proportional to $W_{tree}$. Hence the partition function of the matrix
model is \cite{ow1}
\begin{equation}
Z = e^{-{\cal F}} = 
\int {\cal D}\Phi {\cal D}Q \, \,exp\left\{-\frac{1}{g_s}\left[W(\Phi)+
\sum_{I=1}^{N_f}\left(\tilde{Q} \Phi Q + m_I \tilde{Q}Q\right)\right]\right\}\, ,
\label {pf1}
\end{equation}
where $\Phi$ is $M\times M$ real antisymmetric matrix and $Q,\, \tilde{Q}$ are
$M$ dimensional vectors. For this theory, the Dijkgraaf-Vafa
conjecture can be generalized \cite {ronne1} to obtain effective superpotential
as a function of glueball field $S= Tr W_{\alpha}W^{\alpha}$:
\begin{equation}
W_{eff} = N_c \frac{\partial {\cal F}_{s^2}}{\partial S}
+ 4 {\cal F}_{RP^2} + {\cal F}_{D^2}\, , \label {weff2}
\end{equation}
where ${\cal F}_{s^2}$ is a free energy of a diagram with topology of
sphere, ${\cal F}_{RP^2}$ is free energy of the diagram with cross-cap one
(topology of $RP^2$) and ${\cal F}_{D^2}$ is the free energy of the
diagram with one boundary (topology of $D^2$).
It is also known that \cite {iho1}
\begin{equation}
{\cal F}_{RP^2}=-\frac{1}{2}
\frac{\partial {\cal F}_{s^2}}{\partial S}\, .
\end{equation}
Using this, $W_{eff}$ becomes
\begin{eqnarray}
W_{eff} &=& (N_c-2) \frac{\partial {\cal F}_{s^2}}{\partial S}
+ {\cal F}_{D^2} \nonumber \\
&=& (N_c-2) \frac{\partial {\cal F}_{\chi=2}}{\partial S} + {\cal F}_{\chi=1}\nonumber \\
&=& W_{VY} + (N_c-2) \frac{\partial {\cal F}^{pert}_{\chi=2}}{\partial S} 
+ {\cal F}_{\chi=1}\,,
\label {weff3}
\end{eqnarray}
where $W_{VY}$ denotes the Veneziano-Yankielowicz potential \cite {vy1}.
Here we have absorbed ${\cal F}_{RP^2}$ in ${\cal F}_{\chi=2}$ and
${\cal F}_{\chi=1}$ contains the contribution to the free energy
coming from the fundamental matter.
As proposed by Dijkgraaf-Vafa, we need to take the planar limit of the
matrix model. The planar limit can be obtained by taking 
$(M,N_f \rightarrow \infty)$ as well as $g_s \rightarrow 0 $ such that
$S=g_s M$ and $S_f=g_s N_f$ are finite. 
The total free energy of this matrix model can be expressed as an expansion
in genus $g$ and the number of quark loops $h$: 
\begin{equation} 
{\cal F} = \sum_{g,h} g_s^{2g-2} {S_f}^h {\cal F}_{g,h}(S)~. \label {fe1}
\end{equation} 
Assuming that the fundamental quarks are massive compared to adjoint
matter, we can integrate out the fundamental matter 
fields appearing quadratically in the partition function (\ref{pf1})
to give 
\begin{equation}
Z=e^{-{\cal F}} = \int {\cal D}\Phi \, exp\left[-\frac{1}{g_s}
Tr\left( W(\Phi) + S_f \sum_{I=1}^{N_f} log(\Phi + m_I)\right)\right]~.
\label{fe2}
\end{equation} 
We are now in a position to calculate ${\cal F}_{\chi=1}$ and 
${\cal F}_{\chi=2}$ contributions to the superpotential.
In the following section we apply the 
method developed in \cite {reino1} for the $SO(N_c)$ gauge theory with 
one adjoint matter field and $N_f$ fundamental flavors.

\section{Effective Superpotential}
We wish to compute the exact effective superpotential 
of ${\cal N}=2 $ supersymmetric $SO(N_c)$ gauge theory 
with $N_f$ flavors of quark loops in the fundamental
representation, broken to ${\cal N}=1 $ by addition of a tree level
superpotential $W(\Phi)$ given by eqn.(\ref{tree1}). The supersymmetric
vacua of the theory with superpotential (\ref{tree1}) are obtained by
diagonalizing $\Phi$ such that the eigenvalues are in the set of critical
points of $W(\Phi)$ which are given by the zeros of
\begin{equation} 
W'(x) = g_{2n+2}~ x \prod_{i=1}^n (x^2+a_i^2).
\end{equation} 
Choosing all the $N$ eigenvalues of $\langle \Phi \rangle =0$ gives unbroken
gauge group $SO(N)$. A generic gauge group $SO(N_0)\times \prod_{i=1}^n U(N_i)$
corresponds to $N_0$ eigenvalues of $\langle \Phi \rangle =0$, $N_1$
eigenvalues of $\langle \Phi \rangle =ia_1, \ldots$. We will now look at the
effective superpotential computation for both unbroken and broken gauge group
in the next two subsections.

\subsection{Unbroken gauge Group}
Following the arguments in \cite {reino1}, for the effective 
superpotential evaluation we can still look at a point in the quantum moduli 
space of ${\cal N}=2$ pure gauge theory where 
$r=[N_c/2]$ (rank of $SO(N_c)$) monopoles become 
massless \cite {jo1}. This corresponds to the point 
where the Seiberg-Witten curve factorizes completely.

We have seen that the effective superpotential
of this theory gets contributions from free energies ${\cal F}_{\chi=2}$ and 
${\cal F}_{\chi=1}$ of the matrix model described in the previous section.
We shall first calculate
the contribution to the superpotential coming from ${\cal F}_{\chi=2}$ 
using the moduli associated with Seiberg-Witten factorization.

\subsubsection{Contribution of ${\cal F}_{\chi=2}$}
In this subsection we compute the contribution of ${\cal F}_{\chi=2}$ to the
effective superpotential. This contains free energies of the diagrams having
topology of $S^2$ and $RP^2$. 
Taking derivative of eqn.(\ref{fe1}) with respect to $g_s$ 
\begin{equation} 
\frac{\partial{\cal F}}{\partial g_s} = \sum_{g,h} g_s^{2g-3}(S_f)^h \left(
(2g-2){\cal F}_{g,h} + S \frac{\partial {\cal F}_{g,h}}{\partial S} \right)
+\sum_{g,h} h g_s^{2g-3} (S_f)^h {\cal F}_{g,h}(S)\, . \label{dgsfe}
\end{equation}
According to Dijkgraaf-Vafa, one should take the planar limit on the
matrix model side. Also we take the number of quark loops, $h=0$
for $\chi=2$ free-energy computation.
Planar limit of the above equation gives
\begin{equation} 
\frac{\partial{\cal F}}{\partial g_s} = g_s^{-3} \left(S \frac{\partial {\cal F}_{\chi=2}}
{\partial S} - 2 {\cal F}_{\chi=2}\right)\, .
\end{equation} 
We can also differentiate eqn.(\ref{fe2}) with respect to $g_s$ to give
\begin{equation} 
\frac{\partial{\cal F}}{\partial g_s} = -g_s^{-2} \langle TrW(\Phi) \rangle\, .
\end{equation}
>From the above two equations 
\begin{equation} 
g_s \langle TrW(\Phi) \rangle = 2 {\cal F}_{\chi=2} 
- S \frac{\partial {\cal F}_{\chi=2}}{\partial S}\, . \label{f21}
\end{equation} 
The form of $W(\Phi)$ shows that the LHS contains the vacuum expectation
values $\langle Tr\Phi^{2p} \rangle$. It is clear from the above equation
that once we obtain the vevs $\langle Tr\Phi^{2p} \rangle$, we can easily
compute ${\cal F}_{\chi=2}$. In the case of ${\cal N}=2$ $SO(N_c)$ gauge
theory, the moduli are given by $u_{2p}=\frac{1}{2p}Tr\Phi^{2p}$. 
We are interested in the complete factorization
of the Seiberg-Witten curve. 
The moduli that factorizes the Seiberg-Witten curve are given by \cite{jo1}
\begin{equation} 
\langle u_{2p} \rangle  
 = \frac{N_c-2}{2p} C^p_{2p} \Lambda^{2p}\, ,
\end{equation}
where $C^i_j=\frac{j!}{i!(j-i)!}$ and
$\Lambda$ is the scale governing the running of the gauge coupling
constant. The matrix model calculation of the vevs of the moduli
done in the context of $SU(N_c)$ \cite {nsw1, nsw2} can be
extended to $SO(N_c)$ giving
\begin{equation}
\langle u_{2p} \rangle = (N_c-2) \frac{\partial}{\partial S} \frac{g_s}{2p}
\langle Tr {\Phi}^{2p} \rangle\, .
\end{equation}
It is obvious from the above two equations that
\begin{equation}
\frac{\partial}{\partial S} g_s \langle Tr {\Phi}^{2p} \rangle 
= C^p_{2p} \Lambda^{2p}~. \label{dtp1}
\end{equation} 
We denote the effective superpotential of pure $SO(N_c)$ gauge theory 
by $W_{eff}^0$. From eqn.(\ref{weff3}) we can write
\begin{equation} 
W_{eff}^0 = (N_c-2) \frac{\partial {\cal F}_{\chi=2}}{\partial S} = 
\frac{(N_c-2)}{2} S \left( -log \frac{S}{{\tilde \Lambda}^3} 
+1\right) + (N_c-2) \frac{\partial {\cal F}_{\chi=2}^{pert}}{\partial S}\, .
\label{weff4}
\end{equation} 
The first term in the above equation is the Veneziano-Yankeilowicz 
superpotential \cite {ahn1} and ${\tilde \Lambda}^{3(N_c-2)}$ is the
strong coupling scale of the ${\cal N}=1$ theory. The second term is 
perturbative in glueball superfield $S$ with
\begin{equation}
{\cal F}_{\chi=2}^{pert} = \sum_{n\geq 1} f_n^{\chi=2}(g_{2p}) S^{n+2}\, .
\label{fe3}
\end{equation}
Once we compute the functions $f_n^{\chi=2}(g_{2p})$, we will have
the effective superpotential of $SO(N_c)$ pure gauge theory. In order to
compute these functions we need to take the derivative of (\ref{f21})
with respect to $S$
\begin{equation}
\frac{\partial}{\partial S} g_s \langle TrW(\Phi)\rangle 
= \frac{\partial {\cal F}_{\chi=2}}
{\partial S} - S \frac{\partial^2 {\cal F}_{\chi=2}}{\partial S^2}\, .
\end{equation}
Substituting eqns.(\ref{weff4},\ref{fe3}) in the above equation, we get 
\begin{eqnarray}
(N_c-2) \frac{\partial}{\partial S} g_s \langle TrW(\Phi)\rangle &=& W_{eff}^0 
-S \frac{\partial W_{eff}^0}{\partial S} \nonumber \\\label{dtwp1}
&=& (N_c-2)\left[\frac{S}{2} 
- \sum_{n\geq 1} n(n+2) f_n^{\chi=2}(g_{2p}) S^{n+1}\right]\,. 
\end{eqnarray}

At the critical point of the superpotential, that is when 
$\partial W^0_{eff}/\partial S = 0$, we have
\begin{equation} 
W_{eff}^0 = (N_c-2) \frac{\partial}{\partial S} g_s 
\langle Tr W(\Phi) \rangle 
= \sum_p g_{2p} \langle u_{2p} \rangle \,.
\end{equation}
The glueball superfield can be obtained at the critical point by the 
following relation \cite{jo1}:
\begin{equation}
S=\frac{\partial W_{eff}^0 }{\partial log \Lambda^{N_c-2}} 
= \sum_{p\geq 1} g_{2p} C^p_{2p} \Lambda^{2p}\,. \label{s1}
\end{equation}
Inserting eqn.(\ref{dtp1}) and eqn.(\ref{s1}) in eqn.(\ref{dtwp1}) one gets
\begin{equation}
\sum_{p\geq 1} \frac{1}{2p}g_{2p} C^p_{2p} \Lambda^{2p}
= \frac{1}{2} \sum_{p\geq 1} g_{2p} C^p_{2p} \Lambda^{2p}
-\sum_{n\geq 1} n(n+2) f_n^{\chi=2} (g_{2p}) S^{n+1}\,.
\end{equation}
Substituting glueball field $S$ in terms of $\Lambda$ (\ref {s1}) and
equating the powers of $\Lambda$ on both sides of the above
equation, we can extract the functions $f_n^{\chi=2} (g_{2p})$:  
\begin{eqnarray} 
f_1^{\chi=2} &=& \frac{1}{8} \frac{g_4}{g_2^2} \nonumber \\
f_{n\geq 2}^{\chi=2} &=& \frac{C^{n+1}_{2(n+1)}}{2^{n+2}(n+1)(n+2)}
\frac{g_{2(n+1)}}{g_2^{n+1}} \nonumber \\
&-& \sum_{l=1}^{n-1} \frac{l(l+2)}{n(n+2)}
f_l^{\chi=2} \sum_{{p_1,\ldots p_{l+1}=1} \atop{p_1+\ldots+p_{l+1}=n+1}}^{n+1}
\frac{C^{p_1}_{2p_1}g_{2p}\ldots C^{p_{l+1}}_{2p_{l+1}}g_{2p_{l+1}}}
{2^{n+1}g_2^{n+1}}\,, \label{ff2}
\end{eqnarray}
Now that we have computed the functions $f_n^{\chi=2} (g_{2p})$, the
$\chi=2$ contribution to the
effective superpotential of the $SO(N_c)$ theory with one adjoint
chiral superfield with arbitrary tree level superpotential is known 
exactly. From eqn.(\ref{weff4}) and eqn.(\ref{fe3}), it is given by
\begin{equation} 
W_{eff}^0  = (N_c-2)\left[\frac{S}{2} \left(-log \frac{S}{\tilde\Lambda^3}+1\right)
+ \sum_{n\geq 1}(n+2)f_n^{\chi=2} (g_{2p}) S^{n+1}\right]\,. \label{fw02}
\end{equation} 
In the case of quadratic tree level superpotential, that is when
$g_{2p}=0$ for $p\geq 2$, the functions $f_n^{\chi=2}(g_{2p})$ vanish for all $n$. 
And we can fix the coupling scale ${\tilde \Lambda}^3$ to $2g_2\Lambda^2$ 
by the requirement that $W^0_{eff}$ satisfies equation (\ref{s1}).
The $W_{eff}^0$ for the quartic tree level superpotential can be 
obtained by substituting $g_{2p}=0$ for $p\geq 3$
in the above result (\ref{fw02}): 
\begin{equation}
W_{eff}^0 = W_{VY} + (N_c-2)\left[\frac{3}{2}\left(\frac{g_4}{4g_2^2}\right)S^2
-\frac{9}{2}\left(\frac{g_4^2}{8g_2^4}\right)S^3
+\frac{45}{2}\left(\frac{g_4^3}{16g_2^6}\right)S^4+\ldots\right]\,.\label {fw0i}
\end{equation} 
This is in perfect agreement with the result of \cite{fo1} where it has
been evaluated in terms of the matrix model as well as IIB 
closed string theory on Calabi-Yau with fluxes.
Substitution of $g_{2p}=0$ for $p\geq 4$ in eqn.(\ref{fw02}) gives
$W^0_{eff}$ for the the theory with sixth order potential:
\begin{eqnarray}
W_{eff}^0 &=& W_{VY} + (N_c-2)\left[\frac{3}{8}\frac{g_4}{g_2^2}S^2
+\left(\frac{5}{12}\frac{g_6}{g_2^3} - \frac{9}{16}\frac{g_4^2}{g_2^4}\right)S^3
+\left(-\frac{15}{8}\frac{g_4 g_6}{g_2^5} + \frac{45}{32}\frac{g_4^3}{g_2^6}\right)S^4 \right.\nonumber \\
&+& \left.\ldots \right]~. \label{w06}
\end{eqnarray}
We now compare the result (\ref{fw02}) with the corresponding result in the
$SU(N_c)$ gauge theory with one adjoint matter. The effective 
superpotential of $SU(N_c)$  theory has been obtained in \cite{reino1}. 
Comparison of the effective superpotentials of these two theories provides
the following equivalence:
\begin{equation}
W_{eff}^{0\,SO(N_c)} (g_{2p}) = \frac{N_c-2}{2N_c} \,
W_{eff}^{0\,SU(N_c)} (g^{\prime}_{2p}=2g_{2p})\,,
\end{equation}
which agrees with the relation obtained in \cite {jo1}.
We shall now address the fundamental matter contribution 
${\cal F}_{\chi=1}$ to the effective potential.

\subsubsection{Contribution of ${\cal F}_{\chi=1}$}

We differentiate the free energy given by eqn.(\ref {fe1}) 
with respect to $S_f$
\begin{equation} 
\frac{\partial {\cal F}}{\partial S_f} = \sum_{g,h} h g_s^{2g-2} (S_f)^{h-1}
{\cal F}_{g,h}(S)\,. \label{dfe1}
\end{equation} 
We are interested in genus $g=0$ and one quark loop $h=1$ contribution
in the planar limit $g_s\rightarrow 0$. 
The dominant term from 
eqn.(\ref{dfe1}) is ${\partial}_{S_f} {\cal F} = g_s^{-2} {\cal F}_{\chi=1}$.\\
Differentiation of eqn.(\ref{fe2}) with respect to $S_f$ gives
\begin{equation}
\frac{\partial {\cal F}}{\partial S_f} = g_s^{-1} \sum_{I=1}^{N_f} 
\langle Tr\, log(\Phi+m_I) \rangle\,, \label{dfe2}
\end{equation}  
This implies
\begin{equation}
{\cal F}_{\chi=1} = g_s \sum_{I=1}^{N_f} \langle Tr\, log(\Phi+m_I)\rangle\,.
\label{f11}
\end{equation}
Expanding the above equation around the critical point $\Phi=0$, we get
\begin{equation} 
{\cal F}_{\chi=1} = \sum_{I=1}^{N_f} \left(S\, log\, m_I  
- \sum_{k=1}^{\infty} \frac{(-1)^k}{k m_I^k} g_s\langle Tr{\Phi}^k\rangle\right)\,.
\end{equation}
Differentiating with respect to S and using eqn.(\ref{dtp1}) we get
\begin{equation}
\frac{\partial {\cal F}_{\chi=1}}{\partial S} = \sum_{I=1}^{N_f}
\left(log\, m_I - \sum_{k=1}^{\infty} \frac{1}{2\,k m_I^{2k}} C^k_{2k}\Lambda^{2k}\right)\,.
\end{equation}
Integrating the above equation with respect to $S$ we obtain
\begin{equation}
{\cal F}_{\chi=1} = \sum_{I=1}^{N_f}S\, log\, m_I 
- \sum_{I=1}^{N_f} \sum_{k,l\geq 1}\frac{l g_{2l} C^l_{2l} C^k_{2k}}
{2 k(k+l)m_I^{2k}} \Lambda^{2(k+l)} + D \,,\label{ff1}
\end{equation}
where $D$ is the constant of integration.
We postulate
\begin{equation}
D = \sum_{I=1}^{N_f} W_{tree}(m_I)\,,
\end{equation} 
and we will see in the next section that the result agrees with the one 
obtained from Calabi-Yau geometry with fluxes.
In order to write this expression in powers of $S$, we write 
${\cal F}_{\chi=1}$ as 
\begin{equation} 
{\cal F}_{\chi=1} = S \sum_{I=1}^{N_f} log\, m_I 
+ \sum_{n\geq 1} f_n^{\chi=1} (g_{2p}) S^{n+1} + \sum_{I=1}^{N_f} W_{tree}(m_I)\,.\label{fp1}
\end{equation}
Comparison with eqn.(\ref{ff1}) gives the following recursive relation for
the coefficients $f_n^{\chi=1} (g_{2p})$
\begin{eqnarray} 
f_1^{\chi=1} &=& - \frac{1}{4} \sum_{I=1}^{N_f} \frac{1}{m_I^2 g_2}\nonumber \\
f_{n\geq 1}^{\chi=1} &=& -\frac{1}{2^{n+1} g_2^{n+1}} \left( \sum_{I=1}^{N_f}
\sum_{k,l=1}^n \frac{l g_{2l} C^l_{2l} C^k_{2k}}{2 k(n+1)m_I^{2k}}\right.\nonumber \\
&+&\left. \sum_{q=1}^{n-1} f_q^{\chi=1} \sum_{{p_1,\ldots p_{q+1}} 
\atop{p_1+\ldots+p_{q+1}=n+1}}^{n+1} C^{p_1}_{2p_1} g_{2p_1}\ldots 
C^{p_{q+1}}_{2p_{q+1}} g_{2p_{q+1}} \right)\,. \label{ff12}
\end{eqnarray}
The eqn.(\ref{fp1}) alongwith eqn.(\ref{ff12}) gives the effective 
superpotential from fundamental matter, for the most general $W_{tree}$.
If we substitute $g_{2p}=0$ for $p \geq 2$ in the above result, we get
${\cal F}_{\chi=1}$ for the gauge theory with quadratic superpotential.
It is explicitly given by,
\begin{equation}
{\cal F}_{\chi=1} = \sum_{I=1}^{N_f} \left[S log\, m_I 
-\frac{1}{4} \frac{S^2}{{m_I}^2 g_2} 
- \frac{1}{8} \frac{S^3}{{m_I}^4 {g_2}^2}
- \frac{5}{48} \frac{S^4}{{m_I}^6 {g_2}^3} - \ldots \right] +D \,.\label {qua0}
\end{equation} 
Also substituting $g_{2p}=0$ for $p \geq 3$, we get ${\cal F}_{\chi=1}$ 
for the theory with quartic superpotential.
\begin{eqnarray} 
{\cal F}_{\chi=1} &=& \sum_{I=1}^{N_f} \left[S log\, m_I +
\left(- \frac{1}{4{m_I}^2 g_2}\right)S^2
+ \left(- \frac{1}{8{m_I}^4 {g_2}^2} + \frac{g_4}{4{m_I}^2 {g_2}^3}\right)S^3 \right.\nonumber \\
&+& \left. \left(-\frac{5}{48{m_I}^6 {g_2}^3} + \frac{9g_4}{32{m_I}^4 {g_2}^4}
- \frac{9 g_4^2}{16{m_I}^2 {g_2}^5}\right)S^4 + \ldots \right] + D \,.\label {quar1}
\end{eqnarray}
If we set $g_{2p}=0$ for $p\geq 4$, the resulting theory has sixth order
tree level potential and the corresponding ${\cal F}_{\chi=1}$ is given by
\begin{eqnarray}
{\cal F}_{\chi=1} &=& \sum_{I=1}^{N_f} \left[S log\, m_I +
\left(- \frac{1}{4{m_I}^2 g_2}\right)S^2
+ \left(- \frac{1}{8{m_I}^4 {g_2}^2} + \frac{g_4}{4{m_I}^2 {g_2}^3}\right)S^3 \right.\nonumber \\
&+& \left. \left(-\frac{5}{48{m_I}^6 {g_2}^3} + \frac{9g_4}{32{m_I}^4 {g_2}^4}
- \frac{9 g_4^2}{16{m_I}^2 {g_2}^5} - \frac{15}{16}
\frac{g_6}{m_I^2 g_2^4}\right)S^4 + \ldots \right] + D \,.\label {six1}
\end{eqnarray}
The total effective superpotential of the theory under consideration is
\begin{eqnarray}
W_{eff}&=&W^0_{eff} + {\cal F}_{\chi=1} \nonumber \\
&=& (N_c-2)\left[\frac{S}{2} \left(-log \frac{S}{2 g_2 \Lambda^2}+1\right)
+ \sum_{n\geq 1}(n+2)f_n^{\chi=2} (g_{2p}) S^{n+1}\right] \nonumber \\
&+& S \sum_{I=1}^{N_f} log\, m_I
+ \sum_{n\geq 1} f_n^{\chi=1} (g_{2p}) S^{n+1} + \sum_{I=1}^{N_f} W_{tree}(m_I)
\label{tweff}
\end{eqnarray}

\subsection{Broken Gauge Group}
In the previous subsection, we have computed the effective superpotential
of $SO(N_c)$ supersymmetric gauge theory for unbroken gauge group.
In this section we obtain the effective superpotential for broken gauge group.
In particular we consider the following breaking pattern,
\begin{equation}
SO(N) \rightarrow SO(N_0) \times \prod_{i=1}^n  U(N_i)~,
\end{equation}
such that $N = N_0+2\sum_{i=1}^n N_i$ and for every factor of the gauge group, there
is a glueball superfield $S_i$. We introduce the variables,
$e_0=0,~ e_i=ia_i,~ e_{-i}=-ia_i,~ i=1,2,\ldots,n$. 

\subsubsection{Contribution of ${\cal F}_{\chi=2}$}
Let us first compute the free energy ${\cal F}_{\chi=2}$ for the pure
gauge theory.
For this case the eqn.(\ref{f21}), which has been used to evaluate
${\cal F}_{\chi=2}$ in the case of unbroken gauge group, modifies to
\begin{equation}
g_s \langle TrW(\Phi) \rangle = 2 {\cal F}_{\chi=2} - \sum_{i=-n}^n
S_i \frac{\partial {\cal F}_{\chi=2}}{\partial S_i}~. \label{bf21}
\end{equation}
The free energy ${\cal F}_{\chi=2}$ is a combination of non-perturbative
part, coming from the Veneziano-Yankielowitz term \cite{aip1} and a
perturbative part: 
\begin{equation}
{\cal F}_{\chi=2} = \sum_{i=-n}^n S_i W(e_i)
- \frac{1}{4} \sum_{i=-n}^n S_i^2 
log\left( \frac{S_i}{\alpha \Lambda \Delta_i} \right)
- \frac{1}{2} \sum_{i,j=-n}^n S_i S_j log \left( \frac{e_i-e_j}{\Lambda}\right)
+ \sum_m {\cal F}_{\chi=2}^{(m)}~, \label{bfe2}
\end{equation}
where the perturbative part is contained in ${\cal F}_{\chi=2}^{(m)}$, which
is polynomial of order $m$ in $S_i$.
Substitution of ${\cal F}_{\chi=2}$ given by eqn.(\ref{bfe2})
in eqn.(\ref{bf21}) implies
\begin{equation}
g_s \langle TrW(\Phi) \rangle = \sum_{i=-n}^n W(e_i)S_i +\frac{1}{4} \sum_{i=-n}^n S_i^2
- \sum_{m\geq 3} (m-2) {\cal F}_{\chi=2}^{(m)}~.\label{expect1}
\end{equation}
It is clear from the above equation that, we are close to having the
free energy ${\cal F}_{\chi=2}$ if we can compute the expectation value
$g_s \langle TrW(\Phi) \rangle$.
In order to compute these expectation values, we 
use the following matrix model loop equation \cite{achkr1} :
\begin{equation}
w^2(x) - 2W'(x) w(x) + f_{2n}(x) = 0~,
\end{equation}
where $w(x)$ is a resolvent of the matrix model and $f_{2n}(x)$ is an
even polynomial of order $2n$ which can be chosen to be
\begin{equation}
f_{2n}(x) = 2 W'(x) \sum_{i=-n}^n \frac{\tilde{S}_j}{x-e_j}~,
\end{equation}
where $\tilde{S}_j = \tilde{S}_{-j}$.
The $n+1$ coefficients of the function $f_{2n}(x)$ can be related to
the glueball superfields by computing the following period integral.
\begin{eqnarray}
&S_i& = \frac{1}{2\pi i} \oint_{A_i} w(x) dx = \tilde{S}_i +\nonumber \\
&&\left. \sum_{{p=0}\atop{m\geq2}}^{m} \frac{(2m-3)!!}{p!(m-p)!}
\frac{\tilde{S}^p_i}{(m+p-2)!} \frac{\partial^{m+p-2}}{\partial x^{m+p-2}}
\frac{1}{g_{2n+2}^{m-1} R_i(x)^{m-1}} \left( \sum_{j\neq i} \frac{\tilde{S}_j}{x-e_j}\right)
^{m-p}\right|_{x=e_i} \label{glue1}
\end{eqnarray}
where $A_i$ denote the cycle enclosing the branch point centered in point
$e_i$ of the spectral curve associated with the matrix model 
and $R_i(x)=\prod_{j\ne i} (x-e_j)$. Also note that $S_i=S_{-i}$. 
Using the resolvent $\omega(x)$, the expectation value 
$g_s \langle TrW(\Phi) \rangle$ can be calculated from the
following contour integration:
\begin{eqnarray}
&&g_s \langle TrW(\Phi) \rangle = \frac{1}{2\pi i} \oint_A W(x) w(x) dx
= \sum_{i=-n}^n \tilde{S}_i W(e_i) +\\
&&\left.\sum_{i=-n}^n \sum_{{p=0}\atop{m\geq2}}^{m} \frac{(2m-3)!!}{p!(m-p)!}
\frac{\tilde{S}^p_i}{(m+p-2)!} \frac{\partial^{m+p-2}}{\partial x^{m+p-2}}
\frac{W(x)}{g_{2n+2}^{m-1} R_i(x)^{m-1}} 
\left( \sum_{j\neq i} \frac{\tilde{S}_j}{x-e_j}\right)^{m-p}\right|_{x=e_i}\nonumber 
\end{eqnarray} 
Here the contour $A=\sum_{i=-n}^n A_i$. One can use eqn.(\ref{glue1}) to write the above expression in terms of
$S_i$ instead of $\tilde{S}_i$. The resulting relation can be expressed in
the form of eqn.(\ref{expect1}). And the comparison with eqn.(\ref{expect1})
gives the polynomials ${\cal F}_{\chi=2}^{(m)}$. As an example, for $m=3$
we get
\begin{eqnarray}
g_{2n+2}{\cal F}_{\chi=2}^{(3)} &=& - \frac{1}{2} \sum_{i=-n}^n \sum_{j \neq i}
\sum_{k \neq i} \frac{S_i S_j S_k}{R_i e_{ij} e_{ik}}
+\frac{1}{2} \sum_{i=-n}^n \sum_{j \neq i} \sum_{k \neq i}
\frac{S_i^2 S_j}{R_i e_{ij} e_{ik}} 
+ \frac{1}{4} \sum_{i=-n}^n \sum_{j \neq i} \frac{S_i^2 S_j}{R_i e_{ij}^2}\nonumber \\
&+& \frac{1}{16} \sum_{i=-n}^n \sum_{j \neq i} \sum_{k \neq i}
\frac{S_i^3}{R_i e_{ij} e_{ik}}
- \frac{1}{6} \sum_{i=-n}^n \frac{S_i^3}{R_i}
\left(\sum_{j \neq i} \frac{1}{e_{ij}}\right)^2~,
\end{eqnarray} 
where $e_{ij}=e_i-e_j$ and $R_i=\prod_{j\neq i} (e_i-e_j)$. This result
matches with the one given in \cite{ow1}, where it has been written by using
the relation between free energies of $U(N)$ and $SO(N)$ gauge theories.

\subsubsection{Contribution of ${\cal F}_{\chi=1}$}
For the computation of matter contribution, we incorporate the fact of
broken gauge group in eqn.(\ref{f11}) as follows,
\begin{equation} 
{\cal F}_{\chi=1} = g_s \sum_{I=1}^{N_f} \langle Tr log(\Phi+m_I) \rangle
= \sum_{I=1}^{N_f} \sum_{i=-n}^n S_i log(e_i+m_I)
+ \sum_{m \geq 2} {\cal F}_{\chi=1}^{(m)} \label{bfe11}
\end{equation} 
where ${\cal F}_{\chi=1}^{(m)}$ are polynomials in $S_i$ of order $m$.
We obtain ${\cal F}_{\chi=1}$ by evaluating the expectation value of
$log(x+m_I)$.
\begin{eqnarray}
&&{\cal F}_{\chi=1} 
= \sum_{I=1}^{N_f} \frac{1}{2\pi i} \oint_A log(x+m_I) w(x) dx 
= \sum_{I=1}^{N_f} \sum_{i=-n}^n \tilde{S_i} log(e_i+m_I) + \\
&& \left. \sum_{I=1}^{N_f} \sum_{i=-n}^n \sum_{{p=0}\atop{m\geq2}}^{m} 
\frac{(2m-3)!!}{p!(m-p)!} \frac{\tilde{S}^p_i}{(m+p-2)!} 
\frac{\partial^{m+p-2}}{\partial x^{m+p-2}}
\frac{log(x+m_I)}{g_{2n+2}^{m-1} R_i(x)^{m-1}}
\left( \sum_{j\neq i} \frac{\tilde{S}_j}{x-e_j}\right)^{m-p}\right|_{x=e_i}
\nonumber 
\end{eqnarray}
This result when expressed in terms of $S_i$, can be compared with
eqn.(\ref{bfe11}) to get ${\cal F}_{\chi=1}^{(m)}$.
For $m=2$, the
expression for ${\cal F}_{\chi=1}^{(2)}$ is
\begin{equation}
g_{2n+2} {\cal F}_{\chi=1}^{(2)} = \sum_{I=1}^{N_f} \sum_{i=-n}^n
\left( \frac{S_i}{e_{iI} R_i} \sum_{j\neq i} \frac{S_j}{e_{ij}}
- \frac{S_i^2}{4 e_{iI}^2 R_i}
- \frac{S_i^2 R_i'}{2 e_{iI} R_i^2}\right)
\end{equation} 
where $e_{iI} = e_i + m_I$. This result agrees with \cite{ow1}.
For $m=3$, ${\cal F}_{\chi=1}^{(3)}$ takes the following form:

\begin{eqnarray}
g_{2n+2}^2 {\cal F}_{\chi=1}^{(3)} &=& \sum_{I=1}^{N_f} \sum_{i=-n}^n
\frac{S_i^3}{e_{iI} R_i} \left[-\frac{1}{8 e_{iI}^3 R_i}
-\frac{R_i'}{3 e_{iI}^2 R_i^2} + \frac{R_i''}{8 e_{iI} R_i^2}
-\frac{R_i'^2}{2 e_{iI} R_i^3} - \frac{R_i'''}{6 R_i^2}\right.\nonumber \\
&+&\left. \frac{5}{4} \frac{R_i' R_i''}{R_i^3}
- \frac{3}{2} \frac{R_i'^3}{R_i^4}
-\frac{1}{2} \sum_{j \neq i} \frac{1}{R_j e_{ij}^3}\right]
+ \sum_{I=1}^{N_f} \sum_{i=-n}^n \frac{S_i^2}{e_{iI} R_i}
\left[ \frac{1}{2 e_{iI}^2 R_i}\sum_{j \neq i}\frac{S_j}{e_{ij}}\right.\nonumber \\
&+& \left. \frac{R_i'}{e_{iI} R_i^2}\sum_{j \neq i}\frac{S_j}{e_{ij}}
+\frac{1}{4 e_{iI} R_i}\sum_{j \neq i}\frac{S_j}{e_{ij}^2}
-\frac{5}{4}\frac{R_i''}{R_i^2}\sum_{j \neq i}\frac{S_j}{e_{ij}}
+3\frac{R_i'^2}{R_i^3}\sum_{j \neq i}\frac{S_j}{e_{ij}}\right.\nonumber \\
&+&\left. 2 \frac{R_i'}{R_i^2}\sum_{j \neq i}\frac{S_j}{e_{ij}^2}
-\sum_{j \neq i}\frac{S_j R_j'}{R_j^2 e_{ij}^2}
+\sum_{j \neq i}\frac{S_j}{R_j e_{ij}^3}
+\frac{3}{2}\frac{1}{R_i}\sum_{j \neq i}\frac{S_j}{e_{ij}^3}
+\sum_{{j\neq i}\atop{k\neq i,j}}\frac{S_k}{R_j e_{ij}^2e_{jk}}\right]\nonumber \\
&+&\sum_{I=1}^{N_f} \sum_{i=-n}^n \frac{S_i}{e_{iI} R_i} 
\times\left[\sum_{{j\neq i}\atop{k\neq j}}
\frac{S_j S_k R_j'}{R_j^2 e_{ij} e_{jk}}
+ \sum_{{j\neq i}\atop{k\neq j}}
\frac{S_j S_k}{R_j e_{ij} e_{jk}^2}
-\frac{1}{2} \frac{1}{e_{iI}R_i} \sum_{{j\neq i}\atop{k\neq j}}
\frac{S_j S_k}{e_{ij}e_{ik}} \right.\nonumber \\
&-&\left. \frac{3}{2} \frac{R_i'}{R_i^2}
\sum_{{j\neq i}\atop{k\neq j}} \frac{S_j S_k}{e_{ij}e_{ik}}
-2 \frac{1}{R_i} \sum_{{j\neq i}\atop{k\neq j}} \frac{S_j S_k}{e_{ij}e_{ik}^2}
-\frac{1}{2} \sum_{{j\neq i}\atop{k,l\neq i,j}} \frac{1}{R_j e_{ij}}
\frac{S_k S_l}{e_{jk} e_{jl}} \right.\nonumber \\
&+&\left. \frac{1}{4} \sum_{j\neq i}
\frac{S_j^2 R_j''}{R_j^2 e_{ij}} - \frac{1}{2} \sum_{j\neq i}
\frac{S_j^2 R_j'^2}{R_j^3 e_{ij}}\right] 
\end{eqnarray}

In principle, the above computation can be done to any order $m$.
It is important to realize the power of assimilating
Dijkgraaf-Vafa conjecture and the connections to factorization
of Seiberg-Witten curves which led to 
such precise determination of
$SO(N_c)$ effective superpotential for arbitrary polynomials of tree level
superpotential at generic point in 
the classical moduli space (both unbroken and 
broken gauge group). In order to make sure that the results
are consistent, we need to compare with other approaches.

In the next section, we compare the results with explicit answers  
obtained from geometric approach of dualities.

\section{ Geometric Engineering and Effective $SO$ Superpotential}
We will briefly recapitulate geometric dualities leading to the
computation of $SO$ superpotential.
\subsection{Geometric Transition}
Consider type IIB String theory compactified on an orientifold
of a resolved Calabi-Yau geometry whose singular limit is 
given by eqn.(\ref {singu}). For description of $SO$ gauge group,
$W(x)$ (\ref {singu}) must be even functions of $x$. 
Further, $W'(x)=0$ determines the
eigenvalues of $\Phi$ which can be $0, \pm i a_i's$. 

We are interested in ${\cal N}=1$ $SO(N_c)$ supersymmetric 
gauge theory in four dimensions. This can be realized by 
wrapping $N_c$ D5branes on ${\bf RP}^2$ of the 
orientifolded resolved geometry- i.e., we place all the $N_c$ 
branes at $x=0$ where eigenvalues of $\Phi$ are zero.
Invoking large $N$ duality \cite{vafa1,civ1, eot1}, 
the supersymmetric gauge theory is dual to IIB string 
theory on a deformed Calabi-Yau geometry with fluxes.
The deformed geometry is described by
\begin{equation}
k\equiv W'(x)^2+ f_{2n}(x)+ y^2+z^2 + w^2=0 \,,\label {defm}
\end{equation}
where $f_{2n}(x)$ is a $n$ degree polynomial in $x^2$. 
The three-cycles in this geometry can be given in terms of
basis cycles $A_i,B_i \in H_3(M, {\bf Z})$ ($i=1,2,\ldots 2n+1)$
satisfying symplectic pairing
$$(A_i, B_j)= -(B_j,A_i)= \delta_{ij}~,~~(A_i,A_j)=(B_i,B_j)=0~.$$
Here the pairing $(A,B)$ of three-cycles $A,B$ is defined as the
intersection number. For the deformed Calabi-Yau (\ref {defm}), these
three-cycles are constructed as ${\bf P}^1$ fibration
over the line segments between two critical points 
$x=0^+, 0^-, \pm i a_i^+, \pm i a_i^- \ldots$ 
of $W'(x)^2+f_{2n}(x)$ in $x$-plane. In particular,
$A_0$ cycle corresponds to ${\bf P}^1$ fibration over the
line segment $0^-< x < 0^+$ and $A_i$'s to be fibration over
the line segments $ia_i^-< x < ia_i^+$. The three-cycles
$B_0(B_i's)$  are non-compact and are given by fibrations over line
segments between $0 < x <\Lambda_0 (ia_i^+<x< i \Lambda_0)$
where $\Lambda_0$ is a cut-off. The deformed geometry (\ref {defm})
has ${\bf Z}_2$ symmetry and hence we can restrict the
discussion to the upper half of $x$-plane.
The holomorphic three-form $\Omega$ for the deformed geometry (\ref {defm})
is give by
\begin{equation}
\Omega = 2 {dx \wedge dy \wedge dz \over \partial k/\partial \omega}~.
\end{equation}
The periods $S_i$ and the dual periods $\Pi_i$ for this deformed geometry are 
$$S_i = \int_{A_i} \Omega~,~~ \Pi_i= \int_{B_i} \Omega~.$$
The dual periods in terms of prepotential ${\cal F}(S_i)$ is 
$\Pi_i= \partial {\cal F}/\partial S_i$. Using the fact that these
three cycles can be seen as ${\bf P}^1$ fibrations over appropriate
segments in the $x$-plane, the periods can be rewritten as 
integral over a one-form $\omega$ in the $x$-plane.
That is, $S_0= 1/(2 \pi i) \int_{0^-}^{0^+} \omega~, ~~   
\Pi_0= 1/(2 \pi i) \int_{0^+}^{\Lambda_0} \omega~, \ldots$
where the one-form $\omega$ is given by
\begin{equation}
\omega = 2 dx \left( W'(x)^2+ f_{2n}(x) \right)^{1 \over 2}~.
\end{equation}
The effective superpotential $W_{eff}^0$ (recall the suffix $0$ 
denotes the contribution from adjoint matter field $\Phi$) 
can be obtained as follows
\begin{equation}
-{1 \over 2 \pi i} W_{eff}^0= \int \Omega \wedge (H_R+ \tau H_{NS})\,\label{wef1},
\end{equation}
where $\tau$ is the complexified coupling constant of type IIB strings,
the $H_R$ and $H_{NS}$ denotes the RR-three form and NS-NS
three-form field strengths. 
Inclusion of matter in fundamental representations
in the geometric  framework corresponds to placing D5 branes 
at locations $x= m_a$ where $m_a$'s are the masses of 
$N_f$ fundamental flavors. These locations are not the zeros
of $W'(x)=0$. The fundamental matter contribution 
to the effective potential is given by \cite {ook1}:
\begin{equation}
W_{eff}^{flav}\equiv {\cal F}_{\chi=1}={1 \over 2} \sum_{a=1}^{N_f} 
\int_{m_a}^{\Lambda_0} \omega~. \label {flav}
\end{equation}

For simplicity, we will confine to the classical solution $\Phi=0$ with $N_c$ D5-branes
at $x=0$. The corresponding dual theory will require $RR$-flux  
over $A_0$ cycle alone and a non-zero period $S_0 \equiv S$.
The effective superpotential $W_{eff}^0$ (\ref{wef1}) in terms of 
$\chi = 2$ part of matrix model free energy ${\cal F}(S)$ will be
\begin{equation}
W_{eff}^0 = \left(\frac{N_c}{2}-1\right) \frac{\partial {\cal F}(S)}{\partial S}
= \left(\frac{N_c}{2}-1\right) \int_{0^{+}}^{\Lambda_0} w dx
\end{equation}

It is important to work out explicitly these formal integrals
for specific potentials and compare with our closed form expression
obtained for arbitrary potentials in subsection 3.1.

\subsection{Effective Superpotential for Sixth Order Potential}
In this subsection we consider the ${\cal N}=1$ $SO(N_c)$ gauge theory with
fundamental matter and the following tree level superpotential:
\begin{equation}
W_{tree}(\Phi) = \frac{m}{2}Tr\Phi^2 + \frac{g}{4}Tr\Phi^4
+ \frac{\lambda}{6}Tr\Phi^6\,.
\end{equation} 
The geometry corresponding to this gauge theory is given by
\begin{equation} 
W^{\prime}(x)^2+f_4(x)+y^2+z^2+w^2=0\,,
\end{equation} 
where $f_4(x)$ is an even polynomial of degree 4.\\
We concentrate on the special classical vacuum $\Phi=0$, which 
is sometimes called as one cut solution in the
context of matrix models \cite {fo1}. We require 
the critical points of $W'(x)^2+f_4(x)$ to be $0^+, 0^-$.
This is achieved by the following one form:
\begin{equation} 
\omega = 2\sqrt{W^{\prime}(x)^2+f_4(x)}dx 
= 2\lambda(x^2+a)(x^2+b)\sqrt{x^2-4\mu^2} dx ~, \label{omega}
\end{equation}
where $0^{\pm}= \pm 2 \mu$. Also $a$ and $b$ are related to the couplings of
the tree level potential in the following way,
$$(a+b)=\frac{g}{\lambda} + 2 \mu^2~,$$
$$ab=\frac{m}{\lambda} + 2\frac{g}{\lambda}\mu^2 + 6\mu^4~.$$
The period integral can be computed from
\begin{equation} 
S = \frac{1}{2\pi i} \int_{-2\mu}^{2\mu} \omega dx~.
\end{equation} 
For the sixth order $W_{tree}$, it is explicitly given by
\begin{equation} 
S = 2m\mu^2 + 6g\mu^4 + 20 \lambda \mu^6 ~. \label{sixs}
\end{equation} 
For the given one-form, the $\chi=2$ contribution to the effective
superpotential is
\begin{equation}
W_{eff}^0=\left(\frac{N_c}{2} -1\right){\partial{\cal F} (S) \over \partial S}= 
\left(\frac{N_c}{2}-1\right)
\int_{2 \mu}^{\Lambda_0} \omega dx~.
\end{equation}
After taking the limit $\Lambda_0 \rightarrow  \infty$ and ignoring the
$\Lambda_0$ dependent terms, the above equation leads to
\begin{equation}
W_{eff}^0 = \left(N_c-2\right) \left[S~log(2\mu) - m \mu^2 
- \frac{3}{2} g \mu^4 - \frac{10}{3} \lambda \mu^6 \right]~.
\end{equation}
All higher powers of $\mu$ vanish. Substitution of $S$ from eqn.(\ref{sixs})
in $W_{eff}^0$ obtained from factorization of Sieberg-Witten curve given by
eqn.(\ref{w06}), agrees with the above result.
The effective superpotential that comes from the contribution of flavors 
(\ref {flav}) is
\begin{eqnarray}
W_{eff}^{flavor} &=& -\sum_{I=1}^{N_f} \left[m_I \sqrt{m_I^2-4\mu^2}\left(
\frac{1}{2}m +\frac{1}{4}g m_I^2 +\frac{1}{6}\lambda m_I^4 + \mu^2\left(\frac{g}{2} +\frac{1}{3}\lambda m_I^2 +\lambda \mu^2\right)\right) \right.\nonumber \\
&+& \left.\mu^2\left(m+\frac{3}{2}g\mu^2+\frac{10}{3}\lambda \mu^4\right) + 2\mu^2\left(m + 3g\mu^2 + 10 \lambda \mu^4\right)log(2\Lambda_0)\right. \nonumber \\&-& \left. 2\mu^2\left(m + 3g\mu^2 + 10 \lambda \mu^4\right)log\left(m_I+\sqrt{m_I^2-4\mu^2}\right) \right]~.
\end{eqnarray}
In obtaining the above result, we take the limit $\Lambda_0 \rightarrow \infty$
and ignore the  $\Lambda_0$ dependent terms. Substituting for $S$ in terms
of $\mu^2$ (\ref {sixs}) in eqn. (\ref {six1}), the  result agrees with the
above expression.


Substitution of $\lambda = 0$ in the above equation, we get flavor contribution
of the $SO(N_c)$ gauge theory with quartic tree level superpotential.
\begin{eqnarray}
W_{eff}^{flavor} &=& -\sum_{I=1}^{N_f} \left[m_I \sqrt{m_I^2-4\mu^2}\left(
\frac{1}{2}m +\frac{1}{4}g m_I^2 + \frac{g}{2}\mu^2 \right) 
+ \mu^2\left(m+\frac{3}{2}g\mu^2\right) \right. \\
&+&\left. 2\mu^2\left(m + 3g\mu^2 \right)log(2\Lambda_0)- 2\mu^2\left(m + 3g\mu^2 \right)log\left(m_I+\sqrt{m_I^2-4\mu^2}\right) \right]~.\nonumber
\end{eqnarray}

The corresponding $S$ is given by
\begin{equation} 
S = 2m\mu^2 + 6g\mu^4 ~,
\end{equation} 
which is quadratic in $\mu^2$ and can be solved to give the roots. Discarding
the negative root, we get
\begin{equation} 
\mu^2= -\frac{m}{6g} + \frac{m}{6g} \sqrt{1+\frac{6gS}{m^2}}~.
\end{equation} 
Substituting $\mu^2$ and
rewriting in powers of $S$ agrees with our expansion (\ref {quar1}).
If we take $g \rightarrow 0$ limit in the above equation, we get
$W_{eff}^{flavor}$ of the $SO(N_c)$ gauge theory with quadratic tree level
superpotential. 
{\small
\begin{equation} 
W_{eff}^{flavor} = - \sum_{I=1}^{N_f} \left[\frac{S}{2} + \frac{M {m_I}^2}{2}
\sqrt{1-\frac{2S}{M {m_I}^2}} + S log\left(\frac{\Lambda_0}{m_I}\right)
- S log\left(\frac{1}{2}+\frac{1}{2}\sqrt{1-\frac{2S}{M {m_I}^2}}\right)\right]
\,.
\end{equation}}
If we replace $M$ by $M'/2$, we get the Affleck-Dine-Seiberg
$SU(N)$ superpotential \cite {ads1}. Expanding the above
equation in powers of $S$ agrees with eqn. (\ref {qua0}).

Though we have considered in detail the sixth order potential, it is
straightforward to obtain the effective superpotential for any 
polynomial potential. So far, we have discussed the results for unbroken
gauge group. 

It will be interesting to verify the results in section 3.2 for broken
gauge group also. Some work in this direction has already been reported
in Ref.\cite{fo2} for quartic potential without matter. 
Even though a formal expressions can be written in integral form for 
a general polynomial potential, we still have to work out the results
in a certain limit to compare with the answers in section 3.2. We hope
to report on these aspects in  future.

\section{Summary and Discussion}
In this paper, we have derived $SO(N_c)$ effective superpotential
for the supersymmetric theory with $N_f$ fundamental flavors (\ref{tweff}).
Using Dijkgraaf-Vafa conjecture and also the Sieberg-Witten factorization,
we have obtained the effective superpotential  for a most general
tree level potential $W_{tree}(\Phi^2)$.
We have shown agreement with the results from  the geometric considerations
of superstring dualities for a sixth order tree level polynomial potential. 
We hope to report the explicit computation within geometric framework for the 
broken gauge group in future. 
                                                                                
Though we have concentrated on the $SO$ gauge group, it appears
that the fundamental matter contribution to the $Sp$ (symplectic)
effective superpotential will be identical (${\cal F}_{\chi=1}$).
However, one has to elaborately perform the derivation as done
for $SO$ group. The effective potential in the absence of matter is 
well-studied from various approaches which leads to the replacement of 
factor $N_c-2$ in eqn.(\ref{tweff}) by $N_c+2$ to get $W_{eff}^0$ for $Sp(N_c)$ gauge group.
                                                                                
Within supersymmetric theories, the effective superpotentials for
different regimes like $N_f=N_c$ or $N_f<N_c$ or $N_f > N_c$ could
be addressed \cite{ils1}. Some of these issues have been considered within the
matrix model approach in \cite{br1,afos1}.

We have confined to a specific form of tree level potential which 
breaks ${\cal N}=2$ to ${\cal N}=1$. It will be interesting to 
look at other tree potentials involving more than one adjoint matter.
We hope to report on these issues elsewhere.
\vspace{0.5cm}

\noindent
{\bf Acknowledgments}
                                                                                
\noindent
PB would like to thank CSIR for the grant.
The work of PR is supported by Department of Science and Technology
grant under `` SERC FAST TRACK Scheme for Young Scientists''.

\end{document}